\documentclass[aps,prd,twocolumn,notitlepage,longbibliography,superscriptaddress,nofootinbib,floatfix]{revtex4-2}

\usepackage{amsmath,amssymb,amsfonts}
\usepackage{graphicx}
\usepackage[colorlinks=true,urlcolor=blue,linkcolor=blue,citecolor=blue]{hyperref}

\newcommand{\ket}[1]{|#1\rangle}
\newcommand{\bra}[1]{\langle#1|}

\newcommand{\ZZ}{\mathbb{Z}_2}
\newcommand{\HB}{H_{B}}
\newcommand{\HE}{H_{E}}

\newcommand{\Gtheta}{\ket{G_0(\theta)}}

\begin{document}

\title{Physics-informed quantum algorithms for glueball-like excitations in a $\mathbb{Z}_2$ lattice gauge theory}

\author{Dan-Bo Zhang}
\email{dbzhang@m.scnu.edu.cn}
\affiliation{Key Laboratory of Atomic and Subatomic Structure and Quantum Control (Ministry of Education),  Guangdong Basic Research Center of Excellence for Structure and Fundamental Interactions of Matter, and  School of Physics, South China Normal University, Guangzhou 510006, China}
\affiliation{Guangdong Provincial Key Laboratory of Quantum Engineering and Quantum Materials,  Guangdong-Hong Kong Joint Laboratory of Quantum Matter, and Frontier Research Institute for Physics,\\  South China Normal University, Guangzhou 510006, China}

\begin{abstract}
Glueball spectroscopy and real-time production with quantum computing require three distinct ingredients: a correlated gauge vacuum, a controlled construction of pure-gauge excitations, and a dynamical detector. We develop a physics-informed quantum-algorithm toolbox for these tasks in a $(2+1)$-dimensional $\mathbb{Z}_2$ lattice gauge theory. We use the term \emph{glueball-like} for localized closed-flux excitations on the confining side of this Abelian model, without identifying them with the non-Abelian glueballs of QCD. A loop-gas circuit and Hamiltonian variational ansatz prepare the gauge vacuum, Wilson-loop quantum subspace expansion constructs and characterizes low-lying excitations, and eigenvector continuation imports parameter-dependent dressing without a rapidly enlarged explicit loop basis. A Bethe--Salpeter-type transition amplitude quantifies the spatial broadening of the lightest state. For dynamics, a vacuum-dressed contractible-loop counter measures excess production of localized glueball-like structures. Although demonstrated in an Abelian model, the toolbox separates preparation, construction, compression, characterization, and dynamical detection in a form that is naturally extensible to non-Abelian lattice gauge theories.
\end{abstract}

\maketitle

\section{Introduction}
Glueballs are a characteristic consequence of non-Abelian gauge dynamics. In QCD, gluons themselves carry color charge, so the gauge field can self-interact and form color-singlet states without valence quarks. Early phenomenological discussions of gluonic hadrons appeared in the 1970s~\cite{FritzschMinkowski1975,JaffeJohnson1976,Robson1977}, and the flux-tube picture later provided an intuitive description in which glueballs are associated with closed tubes of gauge flux~\cite{IsgurPaton1985}. More formally, in a pure Yang--Mills theory a glueball appears as an isolated massive excitation with nonzero overlap with gauge-invariant gluonic operators and is classified by its spin, parity, and charge conjugation. The Yang--Mills existence and mass-gap problem concerns the mathematical existence of such a massive pure-gauge quantum field theory; identifying a particular hadron as predominantly gluonic in full QCD is a separate state-identification problem.

Numerical lattice gauge theory has provided the most quantitative evidence for a pure-gauge glueball spectrum. Successive $SU(3)$ lattice calculations established the low-lying spectrum and its continuum extrapolation~\cite{MichaelTeper1989,Bali1993,MorningstarPeardon1999,Chen2006,AthenodorouTeper2020,AthenodorouTeper2021}, while corresponding $SU(N)$ spectra in $2+1$ dimensions provide a closer dimensional benchmark for the present setting~\cite{AthenodorouTeper2017}. Beyond masses, lattice studies have also used Bethe--Salpeter (BS) wave-function amplitudes to characterize the spatial extent of scalar and tensor glueballs~\cite{deForcrandLiu1992,LoanYing2006,Liang2015}. Long before current quantum-computing studies, the three-dimensional $\mathbb Z_2$ gauge theory was used as a simpler confining gauge model: Monte Carlo calculations explicitly analyzed its ``glueball spectrum'' and found support for a flux-tube interpretation~\cite{Caselle1998,Caselle2002}. Experimentally, the main difficulty is mixing between gluonic and quark-antiquark components. BESIII established $J^P=0^{-+}$ for the $X(2370)$ in 2024~\cite{BESIII2024}; a 2026 analysis of the full data set argues that a dominant lightest pseudoscalar-glueball component provides a consistent explanation of its observed properties~\cite{BESIII2026}. This strengthens the experimental case while not removing the general mixing problem~\cite{CredeMeyer2009,Ochs2013}.

Quantum computation provides a complementary route because gauge-theory wave functions and real-time evolution can be represented directly in Hilbert space. Work on lattice gauge theories has progressed from low-dimensional digital simulations and gauge-invariant formulations~\cite{Martinez2016,Zohar2013,ZoharBurrello2015,ZoharDigital2017,Bender2018,Klco2020} to broader roadmaps for high-energy physics~\cite{Bauer2023,DiMeglio2024}. For the specific $(2+1)$D $\mathbb Z_2$ setting, previous quantum studies have addressed spectroscopy without explicit state preparation~\cite{GustafsonLamm2021}, variational and imaginary-time ground-state preparation~\cite{Lumia2022,Sekiyama2026}, resource-efficient or large-scale time-evolution circuits~\cite{Irmejs2023,GustafsonCloud2021}, and charge and string dynamics on superconducting hardware~\cite{Cochran2025}. Glueball-oriented studies now include discrete-subgroup digitizations of $SU(3)$~\cite{Alexandru2022}, variational and subspace calculations in reduced Yang--Mills models~\cite{Butt2023}, and qubit-regularized $SU(2)$ and $SU(3)$ models with massive glueball analogues~\cite{Siew2026SU2,Siew2026SU3}. Alongside these quantum-computing developments, tensor-network simulations have studied string breaking and the formation of disconnected glueball-like loops in the same $(2+1)$D $\mathbb Z_2$ setting~\cite{XuTheory2025}. A recent trapped-ion experiment directly observed gauge-invariant closed-loop excitations in this model~\cite{Xu2026}, while a trapped-ion qudit experiment has accessed genuinely non-Abelian string breaking and gluonic excitations in pure $SU(2)$ dynamics~\cite{John2026}. These experiments naturally access local flux patterns, while a key next step is to turn such bitstring-resolved structures into controlled particle-like observables referenced to an interacting vacuum. These developments motivate an algorithmic question that precedes any platform-specific implementation: which quantum-algorithm ingredients are needed to prepare a gauge vacuum, construct and characterize a controlled class of pure-gauge excitations, and count localized glueball-like excitations in real-time dynamics?

We address this question in a pure $(2+1)$D $\mathbb Z_2$ lattice gauge theory. The model is Abelian and does not contain gluons, so its closed-flux states are not QCD glueballs in a literal sense. Nevertheless, the analogy is physically established on the confining side: earlier high-energy studies of the gauge Ising model treated its massive pure-gauge states as glueballs and related their spectrum to flux-tube or dual bound-state pictures~\cite{Caselle1998,Caselle2002}, while the recent trapped-ion study uses the same closed-loop language dynamically~\cite{Xu2026}. We therefore use \emph{glueball-like excitation} as an operational term for a localized, gauge-invariant closed-flux excitation dressed by quantum fluctuations. By “physics-informed”, we mean that the algorithmic design starts from the physical structure of the problem—gauge constraints, closed-flux excitations, their dressing, symmetry, and spatial structure—and assigns suitable quantum primitives to these tasks, rather than searching a generic many-body Hilbert space with a generic ansatz. Moreover, for dynamics, we complement the spectral construction with a vacuum-dressed counter of simple contractible closed-flux components.

The paper is organized as a quantum-algorithm toolbox around complementary physical tasks. Section~\ref{sec:model} introduces the rotated lattice representation, the confinement--deconfinement structure, and the regime in which a localized closed-flux particle picture is meaningful. Section~\ref{sec:vacuum} prepares the correlated vacuum with a loop-gas circuit followed by a Hamiltonian variational refinement. Section~\ref{sec:qse} constructs low-lying glueball-like states in a Wilson-loop quantum-subspace basis and resolves their energies, symmetry, and local structure. Section~\ref{sec:ec} uses eigenvector continuation to compress parameter-dependent dressing, while Sec.~\ref{sec:bs} characterizes the spatial extent of the lightest state through a BS-inspired transition amplitude. Section~\ref{sec:quench} develops the vacuum-dressed contractible-loop counter for quench dynamics. Section~\ref{sec:discussion} summarizes the resulting preparation--construction--compression--characterization--detection workflow and its extensions.

\section{$\mathbb{Z}_2$ gauge model and closed-flux excitations}
\label{sec:model}

\begin{figure}[t]
\centering
\includegraphics[width=\linewidth]{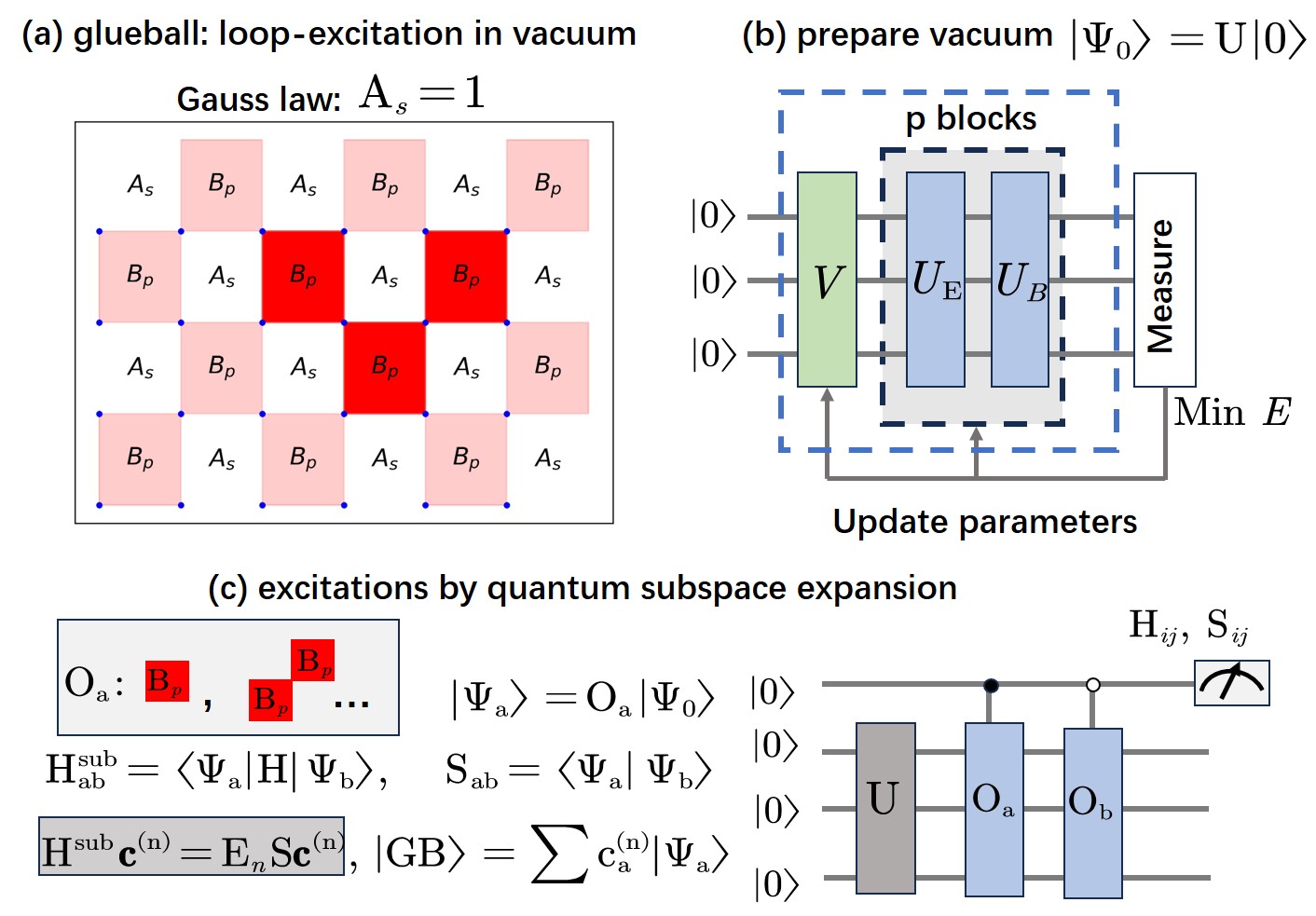}
\caption{Algorithmic workflow. (a) Rotated representation of the $\ZZ$ gauge model. Contractible products of magnetic plaquettes generate closed-flux loops on a vacuum. (b) Vacuum preparation starts from the loop-gas circuit $V(\theta)$ and is refined by gauge-preserving variational layers. (c) Wilson-loop excitations define a nonorthogonal subspace in which low-lying glueball-like states and their observables are obtained from measured overlap and Hamiltonian matrices.}
\label{fig:illustration}
\end{figure}

We use the rotated square-lattice convention shown in Fig.~\ref{fig:illustration}(a). It is obtained by geometrically deforming the usual link representation into a checkerboard in which every four-body star or plaquette acts on the four vertices of one square~\cite{Kitaev2003,Satzinger2021}. The operator structure is closely related to the toric code in a magnetic field, a model with a longstanding phase-diagram literature~\cite{Dusuel2011}; the important distinction here is that Gauss's law is imposed as a kinematic constraint on the physical Hilbert space, so charge-violating toric-code sectors are excluded. This representation is convenient for circuit constructions because the local gauge constraints and magnetic terms appear as uniform four-qubit operators, and closely parallels experimentally used toric-code state-preparation layouts~\cite{Satzinger2021}. The checkerboard squares alternate between magnetic plaquettes $p\in\mathcal P$ and Gauss-law stars $s\in\mathcal S$. For periodic boundary conditions and even $L_x,L_y$, the number of qubits is $N_q=L_xL_y$. We define
\begin{equation}
 B_p=\prod_{i\in p}X_i,\qquad
 A_s=\prod_{i\in s}Z_i,
\label{eq:stabilizers}
\end{equation}
where $X_i$ and $Z_i$ are Pauli operators on gauge qubit $i$. The Hamiltonian is
\begin{equation}
\begin{aligned}
 H(x)&=-(1-x)\HB-x\HE,\\
 \HB&=\sum_{p\in\mathcal P}B_p,\qquad
 \HE=\sum_{i=1}^{N_q}Z_i .
\end{aligned}
\label{eq:H}
\end{equation}
with $0\le x\le1$. Physical states satisfy Gauss's law
\begin{equation}
 A_s\ket{\psi}=\ket{\psi}\quad \text{for every }s\in\mathcal S.
\label{eq:gauss}
\end{equation}
Since $[A_s,H]=0$, the physical sector is preserved exactly by the dynamics and by all variational layers used below.

Two limits make the physics transparent. At $x=1$, Eq.~\eqref{eq:H} reduces to $-\sum_i Z_i$ and the unique vacuum in the chosen topological sector is
\begin{equation}
 \ket{\Omega_E}=\ket{0}^{\otimes N_q},\qquad Z_i\ket{0}=\ket{0}.
\label{eq:electricvac}
\end{equation}
A product of plaquette operators over a connected region $R$,
\begin{equation}
 W_R=\prod_{p\in R}B_p,
\label{eq:WR}
\end{equation}
flips each interior qubit twice and therefore reduces to an $X$ string along the boundary $\partial R$. Thus $W_R\ket{\Omega_E}$ is a closed electric-flux loop and automatically obeys Gauss's law. At $x=1$ its excitation energy is
\begin{equation}
 \Delta E_R=2|\partial R|,
\label{eq:bareenergy}
\end{equation}
where $|\partial R|$ is the number of flipped boundary qubits. An elementary four-link plaquette therefore costs $\Delta E=8$. We call such configurations \emph{bare loop motifs}. Away from $x=1$, the magnetic term mixes different loop shapes, and an eigenstate is generally a coherent superposition of them.

This closed-flux picture is the reason for the glueball analogy used here. In the confining phase of the three-dimensional gauge Ising model, Monte Carlo studies explicitly identified a massive glueball spectrum and found that its organization is consistent with flux-tube and dual bound-state descriptions~\cite{Caselle1998,Caselle2002}. More recently, closed gauge-invariant loops produced dynamically in a $(2+1)$D $\mathbb Z_2$ simulator were likewise interpreted as excitations reminiscent of QCD glueballs~\cite{Xu2026}. Our terminology follows this established analogy while retaining the qualifier ``glueball-like'' to distinguish the Abelian model from Yang--Mills theory.

At $x=0$, the physical ground space is specified by $B_p=+1$ and $A_s=+1$. On a torus it is fourfold degenerate, corresponding to two independent noncontractible Wilson-loop quantum numbers. Because the periodic plaquettes obey one global relation, $\prod_{p\in\mathcal P}B_p=I$, only $N_p-1$ plaquette generators are independent. Choosing one topological sector and an independent set $\mathcal P_0=\mathcal P\setminus\{p_\star\}$, one ground state is
\begin{equation}
 \ket{\Omega_B}=2^{-(N_p-1)/2}
 \prod_{p\in\mathcal P_0}(1+B_p)\ket{\Omega_E}.
\label{eq:magvac}
\end{equation}
The other topological sectors are obtained by noncontractible Wilson loops.

In the thermodynamic limit the transition occurs at~\cite{BloteDeng2002}
\begin{equation}
 \frac{1-x_c}{x_c}\simeq3.04438,
 \qquad x_c\simeq0.2473.
\label{eq:xc}
\end{equation}
The $x<x_c$ side is deconfined, while $x>x_c$ is confined. The distinction is especially transparent in the closed-flux basis. On the confined side, a nonzero electric-string tension penalizes the perimeter of a closed loop, so low-lying pure-gauge excitations can be organized as finite, dressed closed-flux objects. This particle-like picture is clearest deep in the confined regime, close to the electric endpoint $x=1$. As $x$ is lowered toward $x_c$ from above, the tension softens, larger loop configurations acquire increasing weight, and a localized glueball-like state becomes progressively harder to represent with a small-loop basis.

The deconfined side has a qualitatively different interpretation. Its vacuum contains proliferated closed loops; at the exactly solvable point $x=0$, every contractible product $W_R$ is a product of stabilizers and therefore satisfies $W_R\ket{\Omega_B}=\ket{\Omega_B}$. The same closed-loop operator that creates a finite-energy object near $x=1$ is thus part of the vacuum structure rather than an independent particle at $x=0$; in this precise sense there is no separate glueball-like particle associated with $W_R$ at the deconfined endpoint. The elementary toric-code excitations are instead stabilizer violations; within the pure-gauge physical sector fixed by Gauss's law, the relevant local excitations are plaquette-flux violations rather than the contractible closed loops used here. Accordingly, the glueball-state construction developed below is aimed at the confined regime $x>x_c$. The numerical examples at $x=0.4$ and $0.6$ both lie on that side of the thermodynamic transition; $x=0.4$ is used as an intermediate-coupling confined-side benchmark, where substantial finite-size and loop-dressing effects are already visible. The point $x=0.2$ used later is retained only as a demanding benchmark for vacuum preparation~(which can be hard as it is close to the critical point), not as a regime in which we assign a localized glueball interpretation. On the finite $6\times4$ lattice, the loss of a sharply local particle picture can occur well before the thermodynamic transition is reached; the BS-size analysis in Sec.~\ref{sec:bs} makes this finite-size crossover visible.

\section{Gauge-invariant vacuum preparation}
\label{sec:vacuum}

The construction of interacting glueball-like states begins with the quality of the reference vacuum. Rather than starting from a generic hardware-efficient variational circuit, we exploit the gauge structure twice. A loop-gas state first supplies a low-dimensional, gauge-invariant manifold that is exact at the two solvable endpoints and already contains the relevant closed-loop fluctuations. A Hamiltonian variational ansatz then adds correlations using the same electric and magnetic generators that define the model. The first step gives physical initialization; the second provides systematic variational refinement while remaining inside the Gauss-law sector.

\subsection{Loop-gas ansatz and a local preparation circuit}

A natural translation-symmetric interpolation between Eqs.~\eqref{eq:electricvac} and \eqref{eq:magvac}, closely related to the mean-field string-tension ansatz of Ref.~\cite{DusuelVidal2015}, is
\begin{equation}
\begin{aligned}
 \Gtheta&=
 \mathcal N_\theta\prod_{p\in\mathcal P}
 \left(\cos\theta+\sin\theta\,B_p\right)\ket{\Omega_E},\\
 \mathcal N_\theta&=\left[1+(\sin2\theta)^{N_p}\right]^{-1/2},
\end{aligned}
\label{eq:loopgas}
\end{equation}
with $0\le\theta\le\pi/4$. The normalization follows from the single global relation $\prod_{p\in\mathcal P}B_p=I$. It gives $\ket{\Omega_E}$ at $\theta=0$ and $\ket{\Omega_B}$ at $\theta=\pi/4$, and this normalized all-plaquette form is the one used in the numerical results below.

The ansatz also makes its own limitation visible. For a contractible region $R$ containing $|R|$ elementary magnetic plaquettes, one finds on the torus
\begin{equation}
 \langle W_R\rangle_{G_0}
 =\frac{t^{|R|}+t^{N_p-|R|}}{1+t^{N_p}},
 \qquad t=\sin2\theta.
\label{eq:looparea}
\end{equation}
For loops small compared with the system size this reduces to $\langle W_R\rangle\simeq t^{|R|}$ away from the endpoint. Thus, except at $\theta=\pi/4$, the one-parameter state retains the characteristic area-law form and cannot reproduce the generic perimeter-law structure throughout the deconfined phase. Its role is consequently not to be a globally accurate variational family, but to supply a physically informed starting manifold that is exact at both endpoints and already contains coherent closed loops.

The symmetric state in Eq.~\eqref{eq:loopgas} can be prepared with a small extension of the same local stabilizer-state construction used for the toric code~\cite{Kitaev2003,Satzinger2021}. First choose the $N_p-1$ independent magnetic plaquettes $\mathcal P_0=\mathcal P\setminus\{p_\star\}$ in an order such that, before plaquette $p_a$ is processed, one qubit $q_a\in p_a$ is still in $\ket{0}$. The periodic checkerboard requires a suitable ordering and omission of the one dependent plaquette, exactly as in standard toric-code ground-state preparation. For each independent plaquette define
\begin{equation}
 V_a(\theta)=
 \left[\prod_{j\in p_a\setminus\{q_a\}}\mathrm{CNOT}_{q_a\rightarrow j}\right]
 R_y^{(q_a)}(2\theta),
\label{eq:localVa}
\end{equation}
and apply the plaquette blocks sequentially,
\begin{equation}
 V(\theta)=V_{N_p-1}(\theta)\cdots V_2(\theta)V_1(\theta).
\label{eq:Vtheta}
\end{equation}
If the control qubit is still $\ket{0}$ before its block, then for the state $\ket{\varphi}$ of the other plaquette qubits,
\begin{align}
 V_a(\theta)\ket{0}_{q_a}\ket{\varphi}
 &=\cos\theta\ket{0}_{q_a}\ket{\varphi}
 +\sin\theta\ket{1}_{q_a}\!\prod_{j\in p_a\setminus\{q_a\}}X_j\ket{\varphi}\nonumber\\
 &=\left(\cos\theta+\sin\theta B_{p_a}\right)
 \ket{0}_{q_a}\ket{\varphi}.
\label{eq:Vproof}
\end{align}
Iterating this identity prepares the normalized independent-plaquette state
\[
 \ket{\widetilde G_0(\theta)}=
 \prod_{p\in\mathcal P_0}(\cos\theta+\sin\theta B_p)\ket{\Omega_E}.
\]
To restore the omitted dependent factor and hence the translation-symmetric state of Eq.~\eqref{eq:loopgas}, prepare one ancilla in $\cos\theta\ket{0}+\sin\theta\ket{1}$, apply a controlled-$B_{p_\star}$, measure the ancilla in the $X$ basis, and postselect the $+$ outcome. The data register is then proportional to $(\cos\theta+\sin\theta B_{p_\star})\ket{\widetilde G_0(\theta)}$, with success probability
\[
 p_+=\frac{1+(\sin2\theta)^{N_p}}{2}\ge\frac12 .
\]
Thus the all-plaquette symmetry requires only one additional local plaquette operation and a constant-overhead postselection step. At $\theta=\pi/4$, $R_y(\pi/2)$ acting on $\ket{0}$ is equivalent to a Hadamard for this purpose, so Eqs.~\eqref{eq:localVa}--\eqref{eq:Vproof} reduce to the familiar Hadamard-plus-three-CNOT preparation of each independent four-qubit stabilizer.

\subsection{Hamiltonian variational refinement}

The loop-gas state can be systematically refined within the VQE framework~\cite{Peruzzo2014,McClean2016} by a Hamiltonian variational ansatz (HVA), in which variational layers are generated by physically meaningful pieces of the target Hamiltonian. HVA has been studied as a structured many-body ansatz~\cite{Wiersema2020}, used in high-energy calculations of partonic structure~\cite{Li2022}, and applied to critical states of the quantum Rabi model~\cite{Peng2026}; its trainability can also be protected from barren plateaus under controlled parameter conditions~\cite{ParkKilloran2024}. We use
\begin{equation}
 \ket{\Psi_0(\boldsymbol\omega;x)}=
 U_d(\boldsymbol\alpha,\boldsymbol\beta)\Gtheta,
\label{eq:HVAstate}
\end{equation}
with
\begin{equation}
 U_d=\prod_{\nu=1}^{d}
 e^{i\beta_\nu \HB}e^{i\alpha_\nu \HE},
 \qquad
 \boldsymbol\omega=(\theta,\boldsymbol\alpha,\boldsymbol\beta).
\label{eq:HVA}
\end{equation}
Both generators commute with every $A_s$, so gauge invariance is preserved at every layer. The signs in Eq.~\eqref{eq:HVA} are conventional because the angles are variational, while the decomposition into the electric and magnetic generators follows Eq.~\eqref{eq:H}.

\begin{figure*}[t]
\centering
\includegraphics[width=0.95\textwidth]{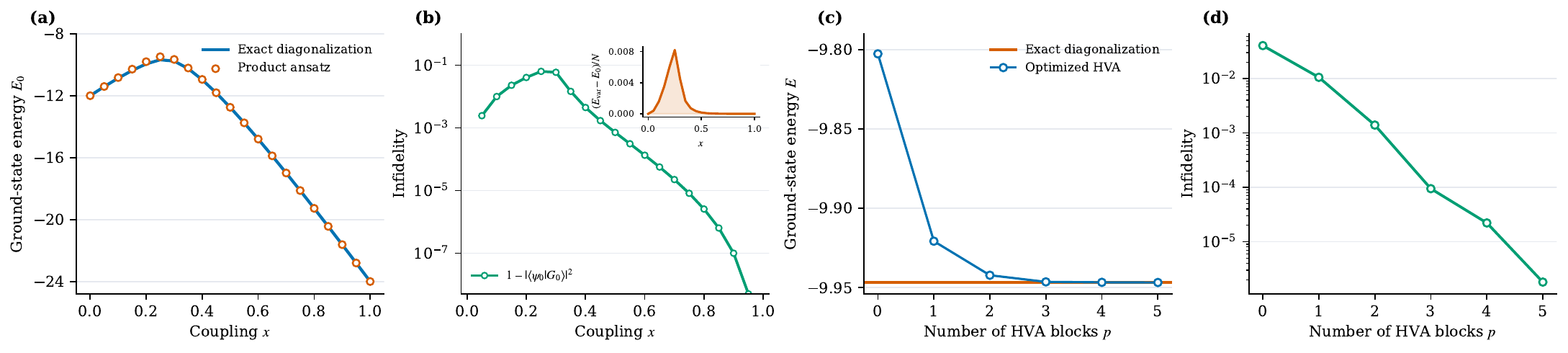}\hfill
\caption{Vacuum preparation on the $6\times4$ periodic lattice. Left~(a and b): comparison of the normalized all-plaquette one-parameter loop-gas ansatz with exact diagonalization over the interpolation parameter $x$. Right(c and d): HVA refinement at the representative point $x=0.2$. The variational energy approaches the exact value and the state infidelity decreases with layer number $p$. The two panels illustrate complementary roles: the loop gas captures the exactly solvable limits and much of the bulk trend, while the HVA repairs the intermediate-coupling correlations that the one-parameter loop gas cannot represent.}
\label{fig:vacuumresults}
\end{figure*}

The parameters are obtained by minimizing
\begin{equation}
 E(\boldsymbol\omega;x)=
 \bra{\Psi_0(\boldsymbol\omega;x)}H(x)\ket{\Psi_0(\boldsymbol\omega;x)}.
\label{eq:energycost}
\end{equation}
Because all terms in $\HB$ mutually commute, and likewise for $\HE$, each HVA layer decomposes into commuting local Pauli rotations. The variational role of $d$ is transparent: $\theta$ establishes the coarse loop density, while alternating electric and magnetic evolutions generate correlations between loops and allow their amplitudes to reorganize beyond the one-parameter loop-gas form of Eq.~\eqref{eq:loopgas}. For the layer-convergence test in Fig.~\ref{fig:vacuumresults} we use $d=0,\ldots,5$ at $x=0.2$; the subsequent HVA vacua use $d=5$. A depth-$d$ ansatz contains $1+2d$ variational parameters. We optimize them with L-BFGS-B using analytic gradients and a multistart strategy combining warm starts from the preceding depth with perturbed and independent random initializations, which reduces sensitivity to local minima in the intermediate-coupling regime.

Figure~\ref{fig:vacuumresults} illustrates this division of labor. The loop-gas ansatz is accurate near the exactly solvable limits and loses fidelity in the intermediate regime, where neither an almost-empty nor an almost-equal-weight loop gas captures the full correlation structure. At $x=0.2$, adding HVA layers rapidly improves both energy and fidelity. The energy can converge more quickly than the fidelity. For the subsequent excited-state calculation, nevertheless, the state fidelity and the quality of correlation functions matter: QSE matrix elements are evaluated on the optimized vacuum, so an apparently small energy error need not guarantee equally small errors in all excited-state observables.

\section{Wilson-loop subspace construction and characterization}
\label{sec:qse}

Once the vacuum has been prepared, the next task is not merely to estimate an excitation gap but to identify the gauge-field structure of the low-lying states. Quantum subspace expansion (QSE) is well suited to this goal because the excitation manifold can be generated by physically interpretable Wilson-loop operators acting on the variational vacuum. Projecting the Hamiltonian into this nonorthogonal manifold yields the low-energy spectrum, while projecting symmetry operators and local loop observables into the same manifold reveals how the states are organized in space and symmetry sectors. We first formulate the loop-based QSE and its truncation, and then apply the resulting subspace to structural observables and the numerical spectrum.

\subsection{Quantum subspace expansion}

We now turn from the vacuum to glueball-like excitations. Let $\{O_a\}_{a=0}^{D-1}$ be a set of gauge-invariant loop operators, with $O_0=I$ and the remaining $O_a=W_{R_a}$ chosen from contractible loop shapes and translations. Acting on the optimized vacuum gives a nonorthogonal basis
\begin{equation}
 \ket{\phi_a}=O_a\ket{\Psi_0}.
\label{eq:qsebasis}
\end{equation}
The projected Schr\"odinger equation is the generalized eigenvalue problem~\cite{McCleanQSE2017,Colless2018}
\begin{equation}
 H^{\rm sub}\mathbf c^{(n)}=E_n S\mathbf c^{(n)},
\label{eq:qse}
\end{equation}
where
\begin{align}
 H^{\rm sub}_{ab}&=\bra{\Psi_0}O_a^\dagger H O_b\ket{\Psi_0},\\
 S_{ab}&=\bra{\Psi_0}O_a^\dagger O_b\ket{\Psi_0}.
\label{eq:qsemat}
\end{align}
The resulting state is
\begin{equation}
 \ket{\Psi_n^{\rm QSE}}=\sum_a c_a^{(n)}O_a\ket{\Psi_0},
 \qquad
 \mathbf c^{(n)\dagger}S\mathbf c^{(n)}=1.
\label{eq:qsestate}
\end{equation}
In the present $\ZZ$ model, every $O_a$ and every term of $H$ is a Pauli string or a product of Pauli strings. Consequently, the projected matrices reduce to expectation values of Pauli strings on $\ket{\Psi_0}$, and commuting strings can be grouped into common measurement settings.

We select loop and training states with sufficient separation to keep the generalized eigenproblem well conditioned without overlap-eigenvalue truncation.

The loop truncation has a physical control parameter. Let $q(W_R)=|R|$ be the number of elementary plaquettes enclosed by a connected loop. If we retain products involving at most $q_{\max}$ independent plaquettes, then
\begin{equation}
 D_{\rm loop}\le
 \sum_{q=0}^{q_{\max}}\binom{N_p-1}{q}
 =O(N_p^{q_{\max}})
\label{eq:loopsizing}
\end{equation}
for fixed $q_{\max}$. Thus the construction is polynomial for a class of low-lying excitations whose wave functions remain concentrated on loops of bounded size. It is not polynomial in general: if the required $q_{\max}$ grows extensively with $N_p$, the basis becomes exponentially large. This is the principal algorithmic failure mode near a regime in which the putative particle is spread over the system.

\subsection{Symmetry and structural observables}

A spectral gap alone does not show what an excitation looks like. QSE makes the wave function accessible in a physically chosen operator basis, but two distinctions are important. First, the coefficients $c_a^{(n)}$ depend on the nonorthogonal basis and are not themselves basis-independent probabilities. Second, exact degeneracies allow arbitrary unitary rotations within the degenerate eigenspace. Structural statements should therefore be tied to symmetry operators and measured observables.

For example, let $\mathcal R$ denote a chosen lattice reflection. In a degenerate QSE eigenspace one can construct
\begin{equation}
 R^{\rm sub}_{ab}=\bra{\phi_a}\mathcal R\ket{\phi_b}
\label{eq:reflectionmat}
\end{equation}
and diagonalize it after the energy problem. The resulting combinations have definite reflection parity $P=\pm1$. Translation operators can be analyzed independently in the same eigenspace when momentum identification is required.

\begin{figure*}[t]
\centering
\includegraphics[width=0.98\textwidth]{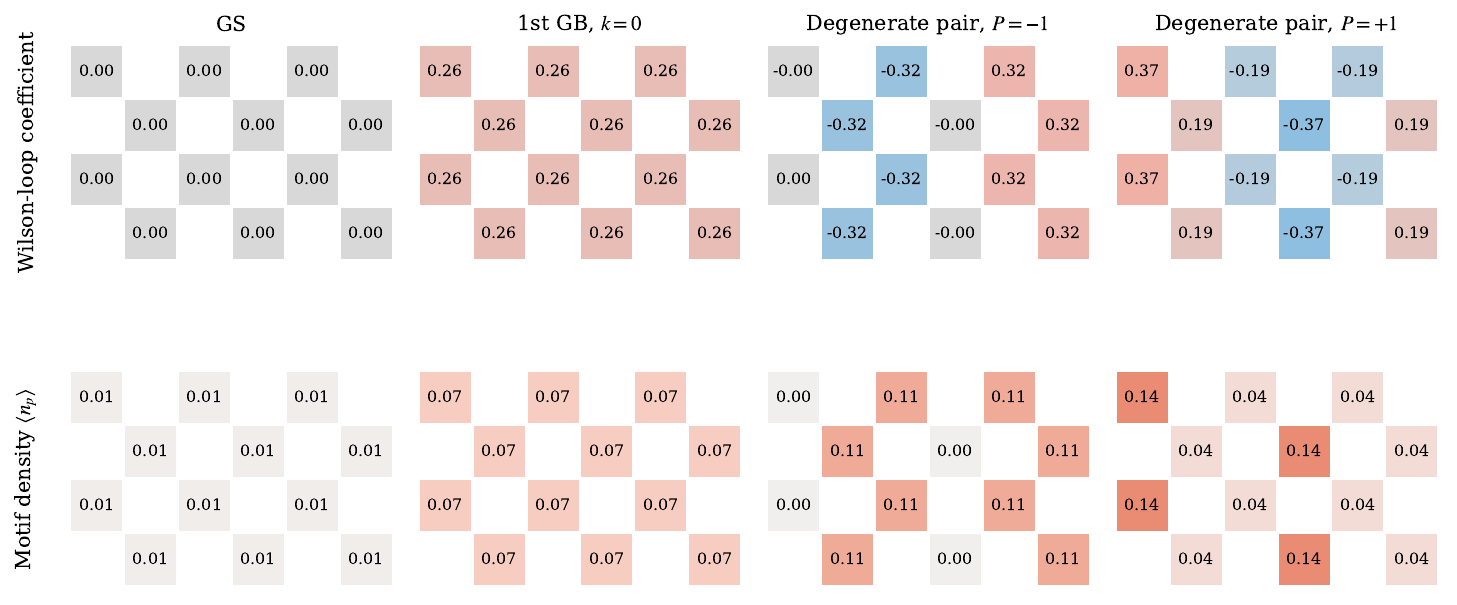}
\caption{Structure of the low-lying states at $x=0.6$. Top: elementary-plaquette Wilson-loop coefficients extracted from a converged QSE basis; the diverging color scale resolves their relative signs. Bottom: local elementary-loop motif density $\langle n_p\rangle$, shown with a sequential color scale. The first excitation is nondegenerate and has $\mathbf k=0$. The following two states form a degenerate finite-momentum pair: diagonalizing the primitive checkerboard translations in this subspace gives eigenvalues $e^{\pm i\pi/3}$. The displayed $P=\pm1$ states are reflection-even/odd standing-wave combinations within this $\pm\mathbf k$ doublet, while the motif density provides a basis-independent view of their spatial support.}
\label{fig:structure}
\end{figure*}

For the local structural plots below we use the elementary-plaquette motif projector
\begin{equation}
 n_p=\prod_{i\in\partial p}\frac{1-Z_i}{2}.
\label{eq:elementarymotif}
\end{equation}
It is the quantity called the ``bare-glueball density'' in Fig.~\ref{fig:structure}. Away from the strong-electric limit it should not be interpreted as a conserved particle-number operator. It is instead a local pattern detector that reveals where the excitation carries elementary closed-flux content; the general loop-motif construction used for nonequilibrium counting is introduced in Sec.~\ref{sec:quench}.

For any observable $O$, QSE requires no reconstructed full wave function. One measures
\begin{equation}
 O^{\rm sub}_{ab}=\bra{\phi_a}O\ket{\phi_b}
\end{equation}
and evaluates
\begin{equation}
 \langle O\rangle_n=\mathbf c^{(n)\dagger}O^{\rm sub}\mathbf c^{(n)}.
\label{eq:qseobs}
\end{equation}
This same recipe gives motif densities, Wilson-loop expectation values, spatial correlators, and symmetry operators.

For the $6\times4$ example at $x=0.6$, we use a converged Wilson-loop QSE basis (the $q_{\max}=6$ construction spans the full reduced sector on this finite torus) and display only the elementary-loop coefficients in Fig.~\ref{fig:structure}. The four low-lying states have vanishing energy variance within the reported numerical precision. The lowest has energy $E_0=-14.8009$. The next state has $E_1=-10.1962$, corresponding to a gap $\Delta_1=4.6047$, while the following pair has $E_{2,3}=-10.1304$ and $\Delta_{2,3}=4.6706$. The isolated first excitation is nondegenerate and diagonalization of the primitive magnetic-lattice translations gives $\mathbf k=0$. The following pair spans a degenerate subspace with translation eigenvalues $e^{\pm i\pi/3}$ for both primitive translations, equivalent to $\mathbf k_{\pm}=(\pm\pi/3,0)$ in the site-coordinate convention. Reflection exchanges the two momentum partners, so diagonalizing reflection within this subspace produces the $P=\pm1$ standing-wave combinations shown in Fig.~\ref{fig:structure}. Thus the parity patterns do not imply two distinct zero-momentum states.

The motif-density panel gives a second view. A uniform or symmetry-patterned $\langle n_p\rangle$ shows how much of the QSE state locally resembles the smallest strong-coupling loop, but it does not by itself define a particle radius. More generally, connected motif correlations remain useful local diagnostics, yet near the transition the interacting vacuum itself contains substantial loop fluctuations and an elementary-motif second moment need not track the spatial extent of the excitation. For this reason, Sec.~\ref{sec:bs} adopts instead a BS-inspired transition amplitude between the interacting vacuum and the lightest excitation, using a symmetry-reduced family of closed-loop probes. The resulting size diagnostic is attached to a physical transition matrix element rather than to the nonorthogonal QSE coefficients or to a particular bare-motif count.

\section{Eigenvector-continuation compression}
\label{sec:ec}

\begin{figure*}[t]
\centering
\includegraphics[width=1.0\textwidth]{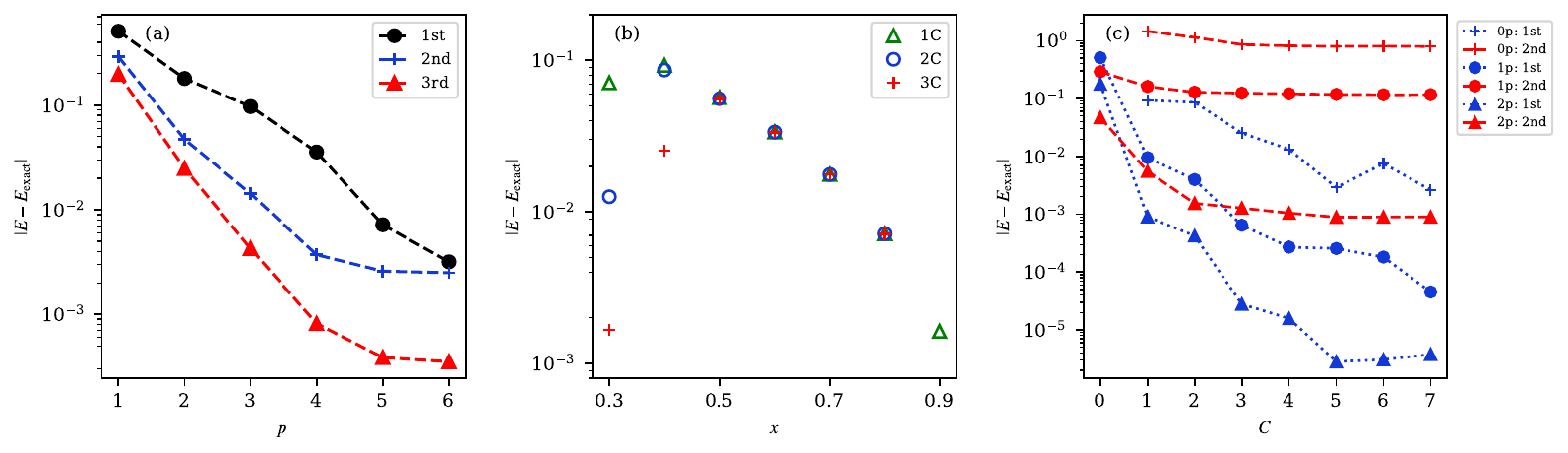}\hfill
\caption{Subspace compression. (a) energy errors of the three lowest glueball-like excitations decrease as the largest Wilson-loop area $n_p$ is increased. (b) vacuum-only EC for the first excitation, using the training set $\{x-0.05C,\ldots,x,\ldots,x+0.05C\}$ with $C=1,2,3$. (c) combined EC and loop bases for the two lowest excitations of the target Hamiltonian at $x=0.4$. The available training vacua are sampled on the grid $x=0.05,0.10,\ldots,0.70$, and $C$ enlarges the selected training set. The labels $0p$, $1p$, and $2p$ denote the three loop-basis choices used in the data, with loop-size cutoffs $n_p=0,1,2$, respectively.}
\label{fig:compression}
\end{figure*}

The loop-only QSE is most economical when the excitation is compact. Figure~\ref{fig:compression}(a) shows the expected trend: enlarging the maximum loop area systematically reduces the energy error of the first three excitations. The improvement also exposes the problem. If accurate states require large $q_{\max}$, Eq.~\eqref{eq:loopsizing} eventually becomes prohibitive.

Nearby ground states provide a complementary low-dimensional direction. Eigenvector continuation exploits the smooth dependence of eigenvectors on Hamiltonian parameters away from singular points~\cite{Frame2018}. For the present interpolation, the relation between two Hamiltonians is particularly simple. For $x,x'\ne0$,
\begin{equation}
 \frac{x}{x'}H(x')=H(x)-\lambda\HB,
 \qquad
 \lambda=\frac{x-x'}{x'}.
\label{eq:scaledH}
\end{equation}
The overall scale $x/x'$ changes eigenvalues but not eigenvectors, so the change from $x$ to a nearby $x'$ can be viewed as a perturbation proportional to $\HB=\sum_p B_p$. This gives a direct physical connection to the loop basis used above. Starting from the vacuum at $x$, perturbation theory constructs the vacuum at $x'$ through successive insertions of the magnetic plaquette operator. A first-order correction contains one plaquette insertion, higher orders contain sequences of several plaquette insertions, and products of such plaquette operators generate closed Wilson-loop configurations of increasing size and complexity. Consequently, a nearby vacuum already contains a coherent mixture of many of the loop sectors that would otherwise have to be introduced explicitly in a single-vacuum QSE basis. In this sense, using vacua at nearby couplings is a compact way of importing the parameter-dependent dressing of the closed-flux configurations.

More formally, to first order, if $Q=I-\ket{\Psi_0}\bra{\Psi_0}$,
\begin{equation}
\begin{aligned}
 \ket{\Psi_0(x')}={}&\ket{\Psi_0(x)}\\
 &+\lambda\,Q[H(x)-E_0]^{-1}Q\HB\ket{\Psi_0(x)}
 +O(\lambda^2),
\end{aligned}
\label{eq:perturbEC}
\end{equation}
up to the usual phase convention. The resolvents in Eq.~\eqref{eq:perturbEC} are important: they dress the plaquette insertions through the dynamics of $H(x)$. Therefore the $q$th-order correction should not be identified with the bare span of loops of area $q$. Rather, the perturbative picture explains why the Wilson-loop basis and the EC basis are complementary. Explicit Wilson loops provide controlled local excitation directions, whereas nearby eigenvectors sample correlated, nonlocal dressing of those directions along the Hamiltonian trajectory.

Choose training points $\{x_\mu\}_{\mu=1}^{N_x}$ and a small set of loop operators $\{O_a\}$. In the numerical implementation used here, the training points are selected at a finite parameter spacing rather than generated as a very dense cluster and subsequently regularized. They are close enough to represent the same smooth low-energy manifold but separated enough to add distinct directions. The combined basis is
\begin{equation}
 \ket{\phi_{\mu a}}=O_a\ket{\Psi_0(x_\mu)}.
\label{eq:ecbasis}
\end{equation}
The composite-index matrices now contain
\begin{align*}
 S_{\mu a,\nu b}&=\bra{\Psi_0(x_\mu)}O_a^\dagger O_b\ket{\Psi_0(x_\nu)},\\
 H_{\mu a,\nu b}&=\bra{\Psi_0(x_\mu)}O_a^\dagger H O_b\ket{\Psi_0(x_\nu)}.
\end{align*}
For $\mu=\nu$ these reduce to the expectation values used in the single-vacuum QSE. For $\mu\neq\nu$, however, they are transition matrix elements between distinct prepared vacua. On quantum hardware they can be measured with a Hadamard-test construction. This introduces additional coherent-depth overhead, although it does not change the dimension of the EC subspace. If $q_{\max}$ is fixed,
\begin{equation}
 D_{\rm EC+loop}\lesssim N_x
 \sum_{q=0}^{q_{\max}}\binom{N_p-1}{q},
\label{eq:ecscaling}
\end{equation}
which remains polynomial when $N_x$ and $q_{\max}$ do not grow exponentially. The gain is that $N_x$ captures smooth nonlocal dressing with parameter, while the local loop operators supply directions that are absent from a vacuum-only manifold.

The three panels of Fig.~\ref{fig:compression} separate these mechanisms. In the loop-only calculation, increasing the maximum loop size from $p=1$ to $p=6$ systematically lowers the energy errors of the first three excitations, demonstrating that larger loops carry real dressing information but also illustrating the rapid growth of the basis. The center panel tests EC without explicit excitation operators. For each target $x$, the subspace contains vacuum states at $x$ and at parameter offsets in steps of $0.05$, with $C=1,2,3$ controlling how many neighboring points are included.  Vacuum-only continuation already performs surprisingly well for the lightest excitation over much of the range in Fig.~\cite{fig:compression}(b), showing that nearby ground states contain useful information about the lowest excited manifold. However, this success is nonuniform and does not remove the need for excitation-oriented basis states when one aims at higher excitations or a more systematic representation.

Fig.~\ref{fig:compression}(c) therefore combines the two resources. For the target Hamiltonian at $x=0.4$, the training pool is sampled at $0.05$ intervals over the displayed range from $x=0.05$ to $0.70$, while $C$ controls the number of selected vacua. With no loop excitation ($0p$), adding vacua alone leaves a sizeable error, especially for the second state. Including the smallest loops ($1p$) reduces both errors substantially, and the enlarged loop set ($2p$, corresponding to the $p=3$ cutoff in the data construction) brings the first excitation to the $10^{-5}$ scale and the second to a few $10^{-3}$ over the larger training sets. Thus the numerical result is not that EC replaces Wilson loops. Rather, the two bases supply complementary information: separated training vacua encode nonlocal parameter-dependent dressing, while explicit Wilson loops provide excitation directions that are weak or absent along the vacuum manifold.

A practical implementation should therefore check the conditioning of $S$ as training states are added and test convergence with both the training set and the loop cutoff. With finite sampling, small overlap eigenvalues can amplify statistical errors in both $H$ and $S$; overlap-eigenvalue thresholding provides a standard stabilization of noisy quantum subspace diagonalization~\cite{Epperly2022}. Recent shadow-based subspace methods provide alternative measurement and regularization strategies~\cite{YangXu2025,Boyd2025}, while bitstring-selected orthogonal configuration subspaces avoid a noisy dense overlap matrix altogether at the price of changing the subspace construction~\cite{Kanno2026}. We do not perform a shot-noise simulation here. In the noiseless data shown, the finite spacing of the chosen training vacua is sufficient for a stable generalized eigenproblem, so overlap truncation is not used. Spatial observables should be converged as well as the energies, since a small energy error alone need not imply a faithful excitation profile. This point is tested directly in the next section by comparing the BS radius obtained from loop-only QSE and from EC-assisted subspaces.

\section{Bethe--Salpeter radius of the lightest excitation}
\label{sec:bs}

The previous sections determine the low-energy states, but a separate observable is needed to ask how spatially extended the lightest glueball-like excitation is. Lattice gauge calculations have long characterized glueball structure through Bethe--Salpeter wave functions of the form $\langle 0|\mathcal O(r)|G\rangle$, where the interacting vacuum and a glueball eigenstate are connected by a probe with a controlled spatial separation~\cite{deForcrandLiu1992,LoanYing2006,Liang2015}. The resulting radius is operator dependent, rather than a unique mass radius, but it provides a direct and established measure of the spatial profile of a chosen glueball interpolating field.

For the present $\ZZ$ model the natural probes are gauge-invariant closed loops, so no gauge fixing is required. We denote the interacting vacuum and the lightest glueball-like state of $H(x)$ by $\ket{\Psi_0(x)}$ and $\ket{\Psi_1(x)}$, respectively. For a connected contractible loop $W_R$, let $\partial R$ contain $N_R$ boundary qubits and define its geometric squared radius on the periodic lattice by the minimum-image form
\begin{equation}
 \rho_R^2=\frac{1}{2N_R^2}
 \sum_{i,j\in\partial R} d_{ij}^2,
\label{eq:loopradius}
\end{equation}
where $d_{ij}$ is the minimum-image distance between boundary qubits. Let $a_p$ denote the nearest-neighbor distance between magnetic-plaquette centers. For one elementary plaquette loop, Eq.~\eqref{eq:loopradius} gives
\begin{equation}
 r_0\equiv\rho_{\square}=\frac{a_p}{2}.
\label{eq:r0}
\end{equation}
We use $r_0$ as the natural length unit below. This normalization makes the strong-electric limit transparent: a state dominated by one elementary closed-flux loop has $R_{\rm BS}/r_0\rightarrow1$.

Translational symmetry and the scalar character of the lightest state allow a substantial reduction of the probe family. We define the dimensionless geometric size $r=\rho_R/r_0$ and group translations, orientations, and symmetry-equivalent connected loops with the same $r$ into
\begin{equation}
 \overline W_r=\frac{1}{\sqrt{N_r}}
 \sum_{R:\rho_R/r_0=r}W_R,
\label{eq:radialloop}
\end{equation}
where $N_r$ is the number of loop operators in the radial class. The discrete BS amplitude is then
\begin{equation}
 \Phi_1(r;x)=
 \bra{\Psi_0(x)}\overline W_r\ket{\Psi_1(x)}.
\label{eq:BSamp}
\end{equation}
More generally, including EC, the same transition matrix element is evaluated on the approximate state, so the observable itself does not depend on how that state was represented.

We characterize the profile by the discrete root-mean-square radius
\begin{equation}
 \left(\frac{R_{\rm BS}(x)}{r_0}\right)^2=
 \frac{\sum_r r^2|\Phi_1(r;x)|^2}
 {\sum_r |\Phi_1(r;x)|^2}.
\label{eq:BSradius}
\end{equation}
This quantity should be interpreted as a BS-inspired effective size associated with the chosen radial loop family. In contrast to the raw QSE coefficients, it is a physical transition amplitude for fixed $\overline W_r$; in contrast to the elementary-motif density, it compares the excitation directly with the interacting vacuum. The absolute value remains operator-family dependent, as in conventional BS wave-function radii, while the evolution of the profile with $x$ directly tracks the transfer of weight from elementary to extended loops. As a robustness check, removing the three largest-$r$ probe classes changes $R_{\rm BS}/r_0$ only from $2.038$ to $2.031$ at $x=0.4$ and from $1.3373$ to $1.3370$ at $x=0.6$; the observed growth is therefore not driven by the largest available loops.

\begin{figure}[t]
\centering
\includegraphics[width=\linewidth]{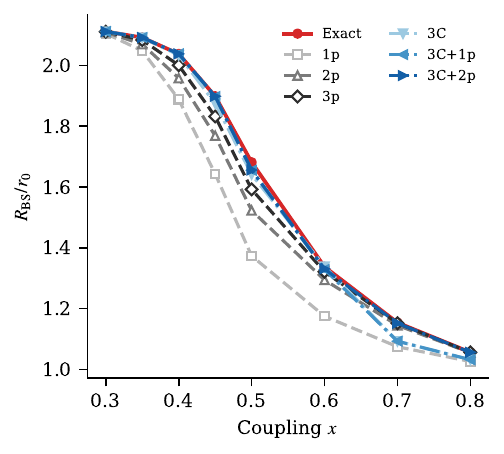}
\caption{Bethe--Salpeter radius of the lightest glueball-like excitation on the $6\times4$ periodic lattice, in units of the elementary-loop radius $r_0=a_p/2$. The red solid curve gives the exact full-basis result. Gray dashed curves with open markers, from light to dark, are loop-only QSE results using five-block HVA vacua initialized from the normalized all-plaquette loop gas; $1p$, $2p$, and $3p$ denote increasing loop cutoffs. Blue dash-dotted curves with filled markers, again from light to dark, use eigenvector continuation with $C=3$: $3C$ contains the seven training vacua only, while $3C+1p$ and $3C+2p$ add explicit loop operators. For every algorithmic curve the bra in Eq.~\eqref{eq:BSamp} is the target HVA vacuum at the same $x$. The strong-electric regime approaches $R_{\rm BS}/r_0\simeq1$, corresponding to an elementary closed-flux loop. Toward smaller $x$, larger loops become important; EC recovers much of this spatial dressing with a smaller explicit loop basis. The flattening near $x\lesssim0.35$ is a finite-size saturation of the $6\times4$ lattice rather than a thermodynamic radius.}
\label{fig:BSradius}
\end{figure}

Figure~\ref{fig:BSradius} shows the lightest-state radius for the same $6\times4$ system. The normalization by $r_0$ gives the strong-coupling result an immediate interpretation. At $x=0.8$ the exact value is $R_{\rm BS}/r_0\simeq1.06$, so the lightest excitation is close to one elementary closed-flux loop, and already the $2p$ loop-only result is essentially converged. As $x$ is lowered, the exact radius grows, reaching approximately $2.04$ at $x=0.4$. The loop-only hierarchy then becomes visibly truncation dependent: at $x=0.4$, the $1p$, $2p$, and $3p$ results give approximately $1.89$, $1.96$, and $2.00$, respectively. This is a structural version of the basis-growth problem seen in Fig.~\ref{fig:compression}: larger loops are not merely correcting the energy but carry the spatial dressing of the excitation.

EC changes this convergence pattern. With $C=3$, the radius at $x=0.4$ is already $R_{\rm BS}/r_0\simeq2.034$ without an explicit loop operator and becomes $2.038$ after adding either $1p$ or $2p$, essentially the exact value $2.038$. Around $x=0.45$--$0.5$ the improvement over a loop-only basis is particularly pronounced. The interpretation is that nearby vacua carry coherent large-loop dressing that would otherwise require a larger explicit Wilson-loop set. Toward $x=0.3$ the exact radius approaches a plateau near $2.11r_0$. Since the largest useful radial probes are already comparable to the geometric scale of the finite $6\times4$ torus, this plateau marks the onset of a finite-size, spatially extended regime rather than a saturation of the thermodynamic glueball radius. Thus $R_{\rm BS}$ serves both as a structural observable and as a practical diagnostic of when the compact glueball-like description underlying a small-loop QSE begins to lose its usefulness.

\section{Dynamical detection and counting of glueball-like excitations}
\label{sec:quench}
The eigenstate constructions above answer which low-lying glueball-like states exist and how they are dressed. A different question arises in real-time dynamics: given a state $\ket{\psi(t)}$, how many localized glueball-like objects are present? This is analogous to counting particles, domain walls, or defects after a quench or a Kibble--Zurek protocol. We initialize the electric vacuum at $x_0=1$ and suddenly change the Hamiltonian to $H(x_1)$,
\begin{equation}
 \ket{\psi(t)}=e^{-iH(x_1)t}\ket{\Omega_E}.
\label{eq:quenchstate}
\end{equation}
Real-time evolution can be implemented by any suitable Hamiltonian-simulation routine; the detector defined below is independent of that choice.

A sum of local motif projectors is not, in general, a particle-number operator. If several $1p$, $2p$, and $3p$ templates overlap on the same physical flux object, the quantity $M_{\rm loop}=\sum_R n_R$ counts several motif occurrences; conversely, a finite motif cutoff can miss a larger loop. We instead define the bare number directly on each computational-basis configuration. For a physical bitstring $z$, let $G_z$ be the graph formed by the occupied $Z_i=-1$ gauge links on the Gauss-law vertices. Gauss's law makes every occupied vertex even-valent. Let $\mathcal C(z)$ denote the connected components of $G_z$, and define
\begin{equation}
 N_{\rm cc}^{(0)}(z)=\sum_{C\in\mathcal C(z)}\chi(C),
\label{eq:Nccclassical}
\end{equation}
where $\chi(C)=1$ when $C$ is a simple contractible closed cycle and $\chi(C)=0$ otherwise. Here contractibility means zero winding around both periodic directions. Components containing degree-four touching or branched vertices are not assigned an unambiguous one-particle interpretation and are excluded from the count; noncontractible winding flux is likewise tracked separately rather than called a glueball-like object. The corresponding diagonal bare number operator is
\begin{equation}
 \widehat N_{\rm cc}^{(0)}
 =\sum_{z\in\mathcal H_{\rm phys}}N_{\rm cc}^{(0)}(z)\ket z\!\bra z.
\label{eq:Nccbare}
\end{equation}
Its integer eigenvalues therefore count distinct localized closed-flux objects without double counting overlapping templates, and arbitrary loop sizes are included automatically by the classical connectivity analysis.

To reference this object count to a correlated confined-side vacuum, we use a common unitary dressing. A direct five-block HVA
\begin{equation}
 D(x_d)=\prod_{\nu=1}^{5}e^{i\beta_\nu^{(d)}H_B}e^{i\alpha_\nu^{(d)}H_E}
\label{eq:Ddressing}
\end{equation}
is optimized so that $D(x_d)\ket{\Omega_E}$ approximates the interacting vacuum at a calibration point $x_d$. We use $x_d=0.6$, where the direct HVA has vacuum infidelity $6.6\times10^{-7}$ on the present lattice. The vacuum-dressed number operator is
\begin{equation}
 \widehat N_{\rm cc}^{(d)}
 =D(x_d)\widehat N_{\rm cc}^{(0)}D^\dagger(x_d).
\label{eq:Nccdressed}
\end{equation}
Unitary conjugation preserves its spectrum, so the same integer counting structure is transported into a vacuum-dressed basis. This is intentionally different from the QSE construction in Sec.~\ref{sec:qse}. The states $W_R\ket{\Psi_0}$ form a generally nonorthogonal variational manifold suited to spectroscopy; weighting motifs by overlaps with a few QSE eigenstates would instead produce a low-energy spectral filter and would discard higher-energy localized objects. For particle counting, we preserve the bare counting algebra and dress the entire number operator by one common unitary.

Operationally, the dressed observable requires no direct decomposition of Eq.~\eqref{eq:Nccdressed} into many Pauli terms. One evolves to $\ket{\psi(t)}$, applies $D^\dagger(x_d)$, performs the same computational-basis measurement used in bitstring-resolved gauge experiments, and classically reconstructs the flux graph and its contractible connected components. Thus
\begin{align}
 N_{\rm cc}^{(0)}(t)&=\bra{\psi(t)}\widehat N_{\rm cc}^{(0)}\ket{\psi(t)},\nonumber\\
 N_{\rm cc}^{(d)}(t)&=\bra{\psi(t)}\widehat N_{\rm cc}^{(d)}\ket{\psi(t)}.
\label{eq:Ncctime}
\end{align}
Because the electric vacuum at $x_0=1$ is not the correlated reference vacuum at $x_d=0.6$, the dressed detector has a small calibration background already at $t=0$. To isolate quench-induced production we therefore plot the excess
\begin{equation}
 \Delta N_{\rm cc}^{(\eta)}(t)
 =N_{\rm cc}^{(\eta)}(t)-N_{\rm cc}^{(\eta)}(0),
 \qquad \eta=0,d.
\label{eq:DeltaNcc}
\end{equation}

\begin{figure}[t]
\centering
\includegraphics[width=\linewidth]{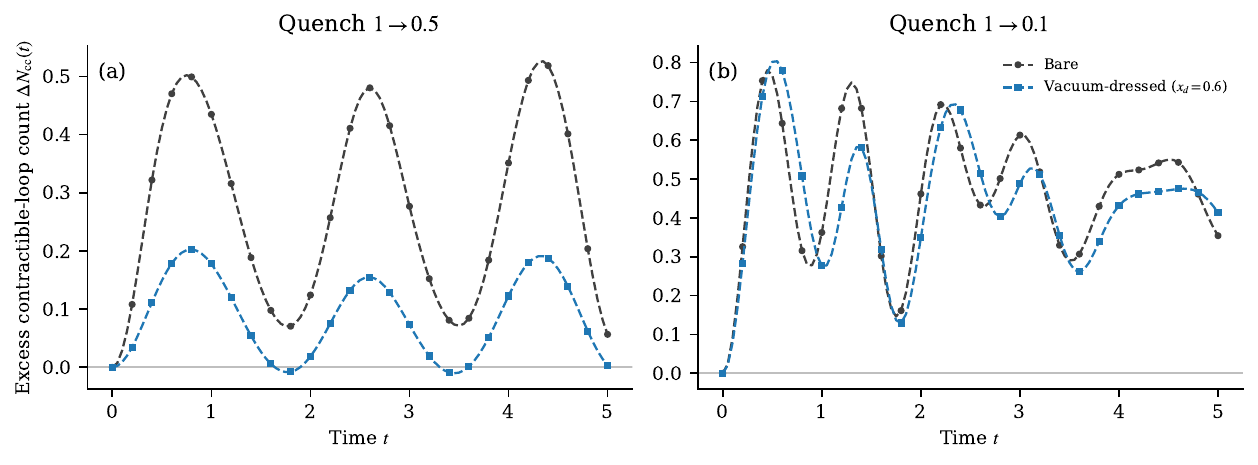}
\caption{Excess production of simple contractible closed-flux components after quenches from the electric vacuum, measured with the bare counter (gray circles) and with the vacuum-dressed counter calibrated at $x_d=0.6$ (blue squares). Both curves are baseline-subtracted according to Eq.~\eqref{eq:DeltaNcc}. (a) Quench to the confined-side point $x_1=0.5$. (b) Deep quench to $x_1=0.1$. The fixed $x_d=0.6$ detector defines the same confined-side glueball-like object in both panels.}
\label{fig:quench}
\end{figure}

Figure~\ref{fig:quench} separates the definition of the detector from the Hamiltonian generating the dynamics. For $x_1=0.5$, the final Hamiltonian remains in the confined regime and the dressed count strongly suppresses the bare closed-loop production, consistent with the fact that only a subset of geometrical flux configurations resemble localized objects relative to the correlated $x_d=0.6$ vacuum. Branched/touching and winding configurations remain rare in this panel. For the deep quench to $x_1=0.1$, the same fixed detector remains well defined, but its interpretation changes: it measures the transient population of confined-side glueball-like local structures inside a state whose natural excitations need not be compact glueballs. In the numerical data, configurations containing degree-four touching/branched flux occur with an average probability of about $25\%$ in the bare dynamics, while noncontractible flux occurs with an average probability of about $28\%$ and can become dominant at individual times. The missing weight therefore provides a direct diagnostic of where the localized-particle description becomes incomplete, rather than a failure of the detector itself.

The present $D$ is trained only on the vacuum, so Eq.~\eqref{eq:Nccdressed} is best described as a \emph{vacuum-dressed} loop-number operator rather than an exact quasiparticle-number operator. A stronger extension would train a common unitary simultaneously on the vacuum and representative one- and multi-glueball states, following the multi-state dressing philosophy of Ref.~\cite{LiFragmentation2024}. QSE can then act as a teacher that identifies the target low-energy manifolds, while the trained unitary compiles that information into a counting basis for subsequent dynamics.

\section{Discussion and outlook}
\label{sec:discussion}

The useful object in this work is not a claim that an Abelian $\ZZ$ theory contains QCD gluons, but a transferable organization of quantum-algorithm primitives around physical questions. The toolbox separates five tasks: the loop-gas circuit and HVA \emph{prepare} a correlated gauge vacuum; Wilson-loop QSE \emph{constructs} the low-energy excitation manifold; EC \emph{compresses} parameter-dependent dressing; symmetry observables and the BS transition amplitude \emph{characterize} the states; and the vacuum-dressed connected-component number \emph{detects and counts} localized glueball-like structures in real time. The BS radius makes the structural step concrete by tracking growth from about one to two elementary-loop scales before finite-size saturation, while the dynamical counter complements QSE spectroscopy by asking how many calibrated local objects occur in an arbitrary nonequilibrium state. Operationally, it requires only inverse dressing followed by bitstring measurement and classical graph analysis.

The regime of efficiency can also be stated sharply. If the low-lying state has appreciable support only on loop operators up to a size $q_{\max}=O(1)$, and if a modest number of EC training points spans its smooth parameter dependence, Eqs.~\eqref{eq:loopsizing} and \eqref{eq:ecscaling} give a polynomial-dimensional generalized eigenproblem. This is the favorable regime demonstrated numerically. The construction can fail near criticality or for highly excited states when the relevant loop size grows with the system and the excitation ceases to be particle-like.  

Several extensions are immediate. Symmetry-projected loop operators can resolve the full finite-lattice irreducible representations and momentum dependence, while larger lattices can determine whether the growth beyond $R_{\rm BS}\sim2r_0$ continues toward the critical point once the finite-size saturation visible here is removed. A complementary, more operator-independent direction would be an energy-density or form-factor radius rather than a BS-wave-function radius. For dynamical detection, a multi-state-trained dressing unitary could transport not only the vacuum but also one- and multi-particle sectors, making $\widehat N_{\rm cc}^{(d)}$ progressively closer to an interacting quasiparticle-number operator. QSE and EC are natural sources of training targets for such a construction, but they need not remain part of the eventual measurement protocol. More ambitiously, the same preparation--construction--characterization--detection architecture can be transferred to non-Abelian digitizations, where Wilson-loop bases acquire representation labels and recoupling structure. Recent $SU(2)$, $SU(3)$, and discrete-subgroup studies~\cite{Alexandru2022,Butt2023,Siew2026SU2,Siew2026SU3,John2026} provide natural test beds. If future non-Abelian simulators can prepare and resolve realistic glueball states, an analogous dressed particle counter could define real-time glueball-production observables after collision-like gauge dynamics. This offers a possible bridge from quantum simulation toward questions motivated by collider glueball searches~\cite{BESIII2024,BESIII2026}, although quantitative contact would additionally require realistic matter content, continuum control, and matching to experimental observables. The present $\ZZ$ model therefore serves as a minimal laboratory in which the full toolbox can be isolated and checked.


\begin{acknowledgments}
We thank Henry Lamm for constructive comments. This work was supported by the National Natural Science Foundation of China (Grant Nos.~12375013 and 12547109) and by the Guangdong Provincial Quantum Science Strategic Initiative (Grant No.~GDZX2503008).
\end{acknowledgments}

\bibliography{glueball_refs}
\end{document}